\documentstyle[11pt,newpasp,twoside,epsf]{article}
\def\edcomment#1{\iffalse\marginpar{\raggedright\sl#1\/}\else\relax\fi}
\reversemarginpar

\begin{document}
\title{Evolution of AGB stars at varying surface C/O ratio: The crucial
effect of molecular opacities}
 \author{Paola Marigo}
\affil{Dipartimento di Astronomia, Universit\`a di Padova, 
                Vicolo dell'Osservatorio 2, 35122 Padova, Italia}

\begin{abstract}
    This study calls attention to the importance of
    properly coupling the molecular opacities to the actual
    surface abundances of TP-AGB stars that experience the third
    dredge-up and/or hot-bottom burning,
    i.e. with surface abundances of carbon and oxygen varying with time. 
    New TP-AGB calculations with variable opacities -- replacing 
    the usually adopted solar-scaled opacity tables -- have proven to
    reproduce, for the first time, basic observables
    of carbon stars, like their effective temperatures, C/O ratios,
    and near-infrared colours.
    Moreover, it turns out that 
    the effect of envelope cooling -- due to the increase in molecular
    opacities -- may cause other important effects, namely:
    i) shortening of the C-star phase; 
    ii) possible reduction or shut-down of the third dredge-up 
    in low-mass carbon stars; and
    iii) weakening or even extinction of hot-bottom burning 
    in intermediate-mass stars. 
\end{abstract}

\section{Introduction}
The observed spectral dichotomy between M-type 
(with C/O$<1$) and C-type (with C/O$>1$) stars
was first explained by Russell (1934) on the basis of 
of molecular equilibria calculations.
It was pointed out that the C/O ratio
plays the key-role in determining the different patterns of molecular 
abundances, hence different dominant molecular bands, characterising the two
classes of giant stars.
 
Despite this basic fact, 
the  description of molecular opacities is still improper
in  most evolution models of AGB stars.
In fact, the usually adopted opacity tables 
(e.g. Alexander \& Ferguson 1994) are strictly valid for 
solar-scaled abundances of elements heavier than helium, 
corresponding to C/O$ = 0.48$ 
(hereinafter also $\kappa_{\rm fix}$ prescription).

Therefore, it is already clear that the inadequacy of 
the opacity prescription
becomes  particularly serious when modelling 
carbon stars, characterised by surface C/O$ > 1$ 
as a consequence of recurrent third dredge-up episodes during the TP-AGB 
evolution. 

Marigo (2002) investigates the effects on the AGB evolution 
due to variable molecular opacities (hereinafter also 
$\kappa_{\rm var}$ prescription), that 
are now computed consistently to the current envelope chemical  
composition of TP-AGB models experiencing the third dredge-up
and hot-bottom burning (HBB).  

As illustrated in the following, 
from comparing the new $\kappa_{\rm var}$ with the 
standard  $\kappa_{\rm fix}$ results, the impact turns out indeed
significant.
\begin{figure}
\plotfiddle{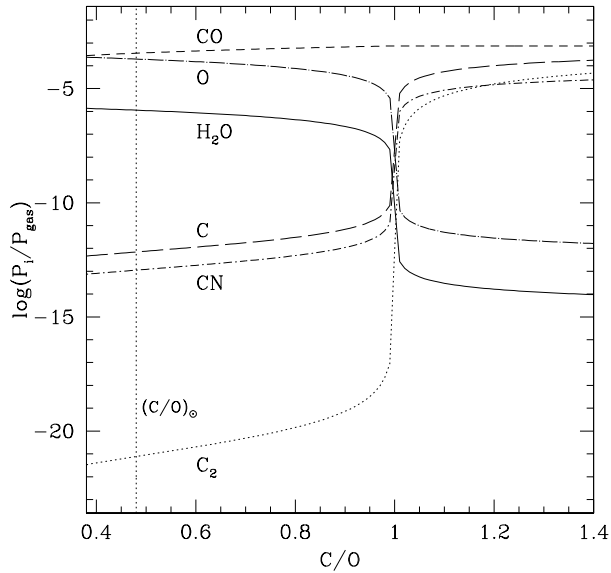}{2.8in}{0.}{42.}{42.}{-135}{-75}
\caption{Molecular abundances (in terms of partial  pressures) 
of a few atomic and molecular species as a function of the C/O ratio,
assuming  a gas pressure $P_{\rm gas} = 10^3$ dyne cm$^{-2}$, and 
a temperature $T = 2500$ K. The vertical line marks  
the molecular concentrations
for a solar composition, with C/O$\sim 0.48$ 
}
\label{fig_molcovar}
\end{figure}
\section{Solar-scaled vs. variable molecular opacities}
A routine has been constructed to derive the molecular 
concentrations through dissociation equilibrium calculations, 
and estimate the opacities due to H$_{2}$, H$_{2}$O, 
OH, C$_{2}$, CN, and CO for any given density, temperature 
and chemical composition of the gas (Marigo 2002).

Figure~1 displays an example of the expected 
chemical pattern at increasing C/O ratio 
(carbon abundance is augmented, while that of oxygen is kept fixed).  
Note the abrupt change in  the chemical abundances,  
as C/O increases from below to above unity, 
for all species but for CO molecule.
In fact, most
atoms of the least abundant element between C and O
are trapped into the CO molecule, due to its high binding energy.

\begin{figure}
\plotfiddle{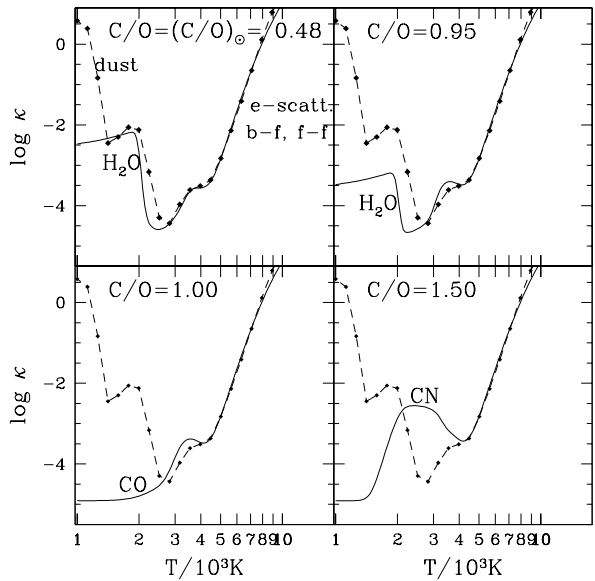}{2.8in}{0.}{42.}{42.}{-135}{-75}
\caption{ Mass absorption coefficient as a function 
of the C/O ratio,
based on Marigo (2002) calculations (solid lines). 
For comparison we also show 
the case of solar-scaled opacities (for C/O$\sim 0.48$) 
according to Alexander \& Ferguson (1994) tables (dashed lines)
}
\label{fig_kcovar}
\end{figure}

Such an abrupt change in molecular abundances at the transition 
from C/O$<1$ to C/O$>1$ results, in turn,  
into a drastic change in the dominant sources of molecular
opacities.
This is illustrated in Fig.~2, based on
Marigo's (2002) calculations (solid line).
As C/O increases the H$_2$O opacity bump, preponderant  
as long as C/O$<1$,  progressively reduces until it completely
disappears for C/O$=1$, and finally the CN opacity bump develops
as soon as C/O$>1$. We also notice that, instead, the solar-scaled opacities
(dashed line) would predict a minimum just in place of the CN bump.

\section{New synthetic TP-AGB models with variable molecular opacities}

Synthetic TP-AGB models, based on complete envelope integrations
(see Marigo et al. 1996, 1998),   
are computed by adopting either the new  $\kappa_{\rm var}$ 
routine, or Alexander \& Ferguson (1994) opacity tables, the standard
$\kappa_{\rm fix}$ prescription. 
A number of important consequences derive
from the comparison of the results.

\subsection{Effective temperatures, C/O ratios, and lifetimes  of C-stars}
The empirical data shown in Fig.~3 clearly indicates 
two major facts, namely: i) the almost 
complete segregation  in effective temperature between 
oxygen-rich and carbon-rich stars; and ii) the relatively low C/O values
($< 2$) measured in carbon-rich stars.

The disagreement is remarkable for 
$\kappa_{\rm fix}$ models, that 
are characterised by too high effective temperatures and C/O ratios.
New $\kappa_{\rm  var}$ models, instead, very well succeed in 
reproducing the observed location  of both oxygen- and carbon-rich
stars in the C/O vs. $T_{\rm eff}$ diagram.

The reason resides just in the abrupt opacity change occurring 
at the transition from the O-rich to C-rich class, which 
causes the large excursion towards lower effective temperatures.
This effect is not present in $\kappa_{\rm fix}$ models.

The photospheric cooling, in turn, favours larger and larger mass-loss rates,
contributing to anticipate the onset of the super-wind, hence the end of the
AGB phase.  
This effect implies a shortening of the C-star phase, and  
a lowering of the typical C/O ratios because of fewer
thermal pulses, hence dredge-up episodes.
We estimate that, for the same model prescriptions (i.e.
dredge-up parameters and mass-loss law), the C-star lifetimes can
be reduced up to a factor of 2 -- 3 when adopting the variable 
opacities.
   
\begin{figure}
\plotfiddle{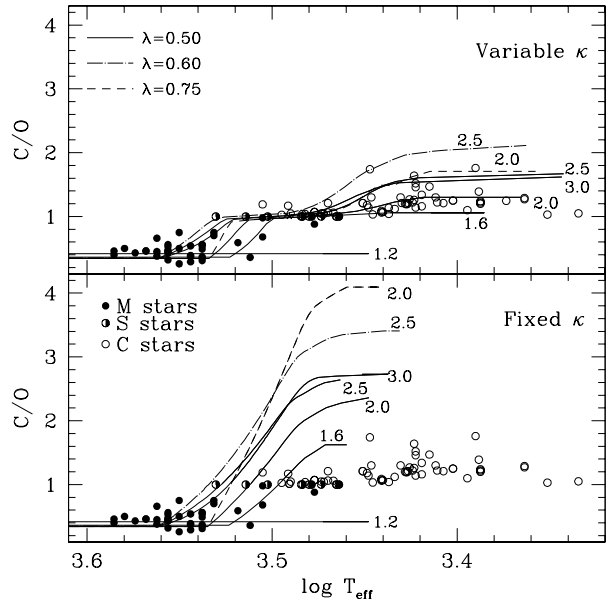}{2.8in}{0.}{42.}{42.}{-135}{-75}
\caption{
Effective temperatures as a function of the C/O ratio
in  Galactic giants.
Abundance determinations are taken from:
Smith \& Lambert (1990 and references therein) for M stars  
(C/O$ <  1$); Ohnaka \& Tsuji (1996) for S stars (C/O$\sim 1$);
Lambert et al. (1986), Ohnaka et al. (2000) for C stars (C/O$ >  1$).
Effective temperatures are taken from the quoted  works,
and Bergeat et al. (2001) for C-stars }
\label{fig_cotef}
\end{figure}

\subsection{Effects on the third dredge-up}
\begin{figure}
\plotfiddle{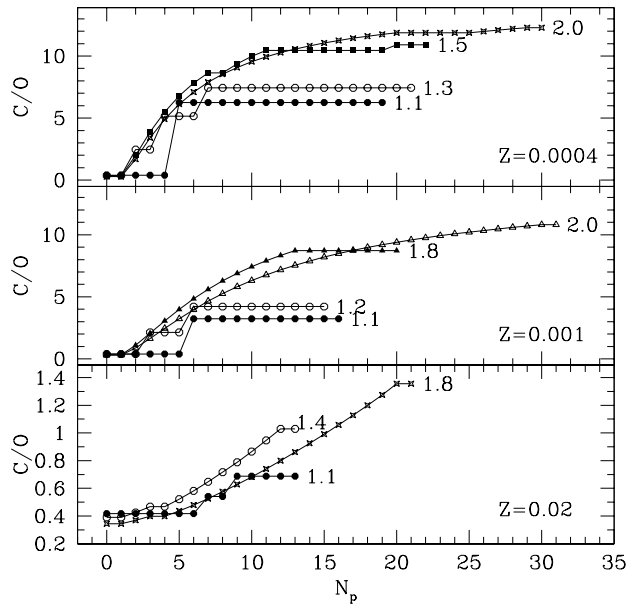}{2.8in}{0.}{42.}{42.}{-135}{-70}
\caption{Evolution of the surface C/O ratio as a function of 
pulse number during the TP-AGB phase of models with different initial
mass and metallicity (as labelled). Calculations are carried out 
with $\kappa_{\rm var}$ prescription and assuming a dredge-up
efficiency  $\lambda=0.5$
 }
\label{fig_npco}
\end{figure}
With the aid of envelope integrations we have explored
the possible effects on the third dredge-up due to
the $\kappa_{\rm var}$ prescription.
In summary it  turns out that, 
at the stage of the post-flash luminosity maximum
(corresponding to the maximum penetration of envelope convection), 
the envelope base temperature, $T_{\rm b}$, 
may significantly decrease as soon as C/O$>1$.

This fact would translate into a likely reduction of the third 
dredge-up efficiency, even determining a possible inhibition of
further dredge-up events.
In other words, besides the reduction of the envelope mass due to mass
loss, another  possible mechanism leading to the  
shut-down of the third dredge-up may be related to 
the envelope cooling caused by opacity effects. 

Preliminary calculations (see Fig.~4) indicate that, especially at lower
metallicities, low-mass TP-AGB stars may experience just one single 
dredge-up episode, become carbon stars, and thereafter preserve
their chemical composition unchanged.
It is also interesting to notice that, in case of solar metallicity 
low-mass stars,
the freezing of the surface abundances as soon as C/O$\ga 1$ could provide
a possible explanation to the formation of S-type stars.
At larger masses the recurrence of dredge-up episodes could  have 
an intermittent behaviour, the 
final shut-down possibly occurring at later stages.
\subsection{Effects on hot-bottom burning}
\begin{figure}[t]
\plotfiddle{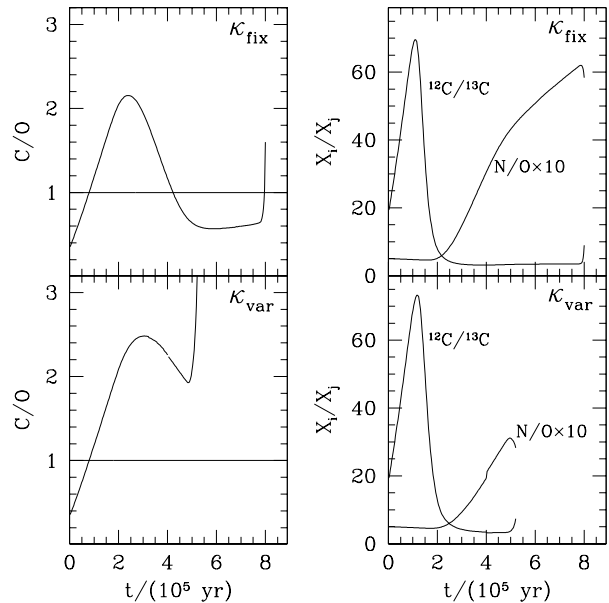}{2.8in}{0.}{42.}{42.}{-135}{-70}
\caption{Evolution of surface abundance ratios
(elemental abundances are expressed by number, in mole g$^{-1}$) 
in the envelope of a (4.5 $M_{\odot}, Z=0.004$) model, 
suffering  both the third dredge-up
and HBB. Calculations are carried out over the entire TP-AGB evolution. 
Note that in the $\kappa_{\rm var}$ model, 
once the transition to carbon star (C/O$>1$) has occurred,  
the C/O ratio no more decreases below unity, contrarily 
to the $\kappa_{\rm fix}$ case }
\label{fig_hbb}
\end{figure}
Our investigation is then extended to stars with larger masses,
say $M > 4.5 M_{\odot}$ (Marigo 2003, in preparation).
It turns out that variable molecular opacities may significantly
influence the structure and nucleosynthesis of massive AGB stars 
by affecting  the efficiency of HBB.
 
Specifically, test calculations with $\kappa_{\rm var}$ 
prescription (see Fig.~5)
indicate that {\sl if } during the early stages  of its TP-AGB evolution, 
a massive  AGB star has experienced efficient carbon dredge-up 
becoming a carbon star, then HBB may be significantly weakened or even 
prevented. The main reason is two-fold: 
The increase in opacities as soon as C/O$>1$ produces a cooling of
both i) the 
envelope base -- hence lowering the nuclear reaction rates --, and
ii) the surface -- hence lowering $T_{\rm eff}$ with consequent 
earlier attainment of larger mass-loss rates.

All these feedback effects are missed in the case of  
$\kappa_{\rm fix}$ models, as we see by comparing the 
evolution of envelope chemical abundances presented in
Fig.~5. For instance, note that at some stage 
the $\kappa_{\rm var}$ model
should appear as a luminous J-type carbon star (with C/O$>1$ and
$^{12}$C/$^{13}$C close to the equilibrium value), whereas 
the $\kappa_{\rm fix}$ model would belong to the class of luminous
M-type giants. 

\section{Conclusions}
This explorative study has shown how large is 
the impact of introducing variable molecular opacities in AGB models.
This has improved the comparison with 
observations and brought  many new results, which may critically 
change and revitalise, in various aspects, 
the present scenario of AGB evolution. 

\acknowledgments
It is a pleasure  to thank G. Meynet, C. Charbonnel, and D. Schaerer
for their kindness and the successful organisation  of the workshop, 
and  Prof. A. Maeder (many wishes!) for fruitful discussions.


\begin{references}
\reference{}
\reference{}
\reference{Alexander D.R., \& Ferguson J.W. 1994, \apj, 437, 879}
\reference{Bergeat J., Knapik A., \& Rutily B. 2001, \aap, 369, 178}
\reference{Lambert D.L., Gustafsson B., Eriksson K., \& Hinkle K.H. 1986,
        \apjs, 62, 373}
\reference{Marigo P. 2002, \aap, 387, 507}
\reference{Marigo P., Bressan A., \& Chiosi C. 1998, \aap, 331, 564}
\reference{Marigo P., Bressan A., \& Chiosi C. 1996, \aap, 313, 545}
\reference{Ohnaka K., Tsuji T., \& Aoki W. 2000, \aap, 353, 528}
\reference{Ohnaka K., \& Tsuji T. 1996, \aap, 310, 933}
\reference{Smith V., \& Lambert D.L. 1990, \apjs 72, 387}
\end{references}
\end{document}